\documentclass[10pt, conference, letterpaper]{IEEEtran}
\makeatletter

\newcommand{\Rmnum}[1]{\expandafter\@slowromancap\romannumeral #1@}
\makeatother
   
\usepackage{cite}
\usepackage{graphicx}
\usepackage{float}
\usepackage{epstopdf}
\usepackage{url}
\usepackage{subfigure}
\usepackage{amsmath}
\usepackage{algorithm}
\usepackage{algpseudocode}
\usepackage{bm}

\usepackage{multirow}
\usepackage{enumerate}
\usepackage{amsthm}

\usepackage[table]{xcolor}

\usepackage[T1]{fontenc}
\usepackage[font=small,labelfont=bf,tableposition=top]{caption}

\DeclareCaptionLabelFormat{andtable}{#1~#2  \&  \tablename~\thetable}

\newtheorem{Lemma}{Lemma}

\newtheorem{Theorem}{Theorem}

\algnewcommand{\Initialize}[1]{%
  \State \textbf{initialize:}
  \Statex \hspace*{\algorithmicindent}\parbox[t]{.8\linewidth}{\raggedright #1}
}

\usepackage{tabularx}

\usepackage{flushend}

\usepackage[T1]{fontenc}
\usepackage[utf8]{inputenc}

\begin{document}

\title{Fair Airtime Allocation for Content Dissemination in WiFi-Direct-Based Mobile Social Networks}

\author{\IEEEauthorblockN{Zhifei Mao and
Yuming Jiang}
\IEEEauthorblockA{Department of Information Security and Communication Technology \\
Norwegian University of Science and Technology (NTNU), Norway} Email: \{zhifeim, jiang\}@item.ntnu.no}

\maketitle

\begin{abstract}
The vast penetration of smart mobile devices provides a unique opportunity to make mobile social networking pervasive by leveraging the feature of short-range wireless communication technologies (e.g. WiFi Direct). In this paper, we study local content dissemination in WiFi-Direct-based mobile social networks (MSNs). We propose a simple GO-coordinated dissemination strategy, as WiFi Direct does not originally support content dissemination. Due to mobility and the short transmission range, the duration of nodes in contact tends to be limited and consequently they compete for the limited airtime to disseminate their own data. Therefore, fair allocation of the limited airtime among the nodes is required. We focus on fairness in content dissemination rate, which is a key application-layer metric, rather than fairness in throughput or airtime and formulate the allocation problem as a generalized Nash bargaining game wherein the nodes bargain for a share of the limited airtime. The game is proved to have a unique optimal solution, and an algorithm with low complexity is designed to find the optimal solution. Furthermore, we propose a detailed scheduling approach to implement the optimal solution. We also present numerical results to evaluate the Nash bargaining based allocation and scheduling. 
\end{abstract}


\IEEEpeerreviewmaketitle

\section{Introduction}

Mobile social networks (MSNs) are new platforms that enable people to share content and form groups without Internet access. By exploiting short-distance wireless communication, people in MSNs can exchange information whenever their devices are within each other's transmission range. To deal with intermittent connectivity due to mobility and short communication range, MSNs employ a store-carry-forward scheme to deliver data. It means that each mobile node may carry different kinds of information for other nodes. Therefore, nodes may need to exchange a large amount of data when they come into each other's range, especially when the MSN is used for multimedia content dissemination and offloading. 

WiFi Direct \cite{alliance2014wi}, which supports typical WiFi speeds and a transmission range up to $200m$, is a favorable technology for data dissemination in MSNs. WiFi Direct devices connect to each other by forming groups. In a group, one of the WiFi Direct devices is selected as group owner (GO) to control the group like a conventional access point (AP), while other nodes connect to the GO as clients. Recently, researchers have demonstrated the feasibility of using WiFi Direct as the medium for opportunistic networking \cite{conti2013experimenting}, multi-hop networking \cite{FunaiTH16}, and multi-group networking \cite{casetti2015content} which are candidate underlying networking techniques for MSNs. In addition, there are already a few WiFi-Direct-based MSN applications that feature content dissemination, such as CAMEO\cite{arnaboldi2014cameo}, public safety\cite{duan2014wi}, and social commerce service\cite{kim2015vada}.


In the literature, there are a plethora of content dissemination protocols for MSNs \cite{zhu2015data,mao2016mobile}. However, most of them do not consider the specifics of underlying mobile networks in their design and ignore problems such as channel allocation and transmission scheduling. In this paper, we focus on local content dissemination within a WiFi Direct group. By its original design, WiFi Direct does not define client to client communication. To allow the data of all nodes being shared with others, we propose a GO-coordinated dissemination strategy where the clients upload their data to the GO that later broadcasts the received data for them. 

Typically, the nodes in a WiFi Direct group cannot exchange as much data as they want since the contact duration can be highly limited due to their mobility. Therefore, a fair allocation of the limited airtime among the nodes is required. The problem of fair airtime allocation in traditional WiFi networks (or WLANs) is a well-studied topic in the literature. Two most studied fairness notions are throughput-based fairness and time-based fairness \cite{tan2004time}, meaning contending nodes obtain equal share of the throughput and airtime respectively\footnote{Throughput-based fairness and time-based fairness can be translated into max-min fairness and proportional fairness, respectively \cite{shi2014fairness,patras2016rigorous}.}. In local content dissemination, however, the meaning of throughput or airtime is not direct to the nodes. Rather, content dissemination rate is a more meaningful metric, as all nodes want to disseminate their data to other nodes in a WiFi Direct group as fast as possible. Therefore, we aim to achieve fairness in content dissemination rate. In fact, equal throughput or airtime does not result in equal dissemination rate. The reason is that the GO has to forward data for the clients, and thus part of its throughput or airtime will be used to disseminate other nodes' data. For the same reason, the node that is selected to be the GO contributes more resources (e.g. battery power and storage) than other nodes. Such asymmetric contributions of nodes are not captured by allocation schemes that achieve throughput-based fairness and time-based fairness.

In this work, we take advantage of a game-theoretic approach, and model the airtime allocation problem as a generalized Nash bargaining game, which yields a unique solution that maximizes social welfare and guarantees fairness in dissemination rate. In summery, we make the following contributions: 1) we propose a GO-coordinated dissemination strategy that enables content dissemination among nodes in a WiFi Direct group; 2) considering the cooperative and self-interested nature of the nodes, we model the airtime allocation as a generalized Nash bargaining game, which captures the asymmetric contributions of nodes, and prove the existence of a unique optimal solution to the game; 3) we present an algorithm with low complexity to find the optimal solution; and 4) to implement the optimal allocation, we design a time-slotted scheduling approach that divides the allocated time into small slots and allows the nodes to transmit data in a round-robin way. 

The rest is organized as follows. Sec. \ref{pro} provides a brief overview of WiFi Direct and introduces the GO-coordinated dissemination. Sec. \ref{GNBSmodel} presents an airtime allocation scheme for the GO-coordinated dissemination using generalized Nash bargaining. A detailed algorithm is designed in Sec. \ref{sec-algorithm}. In Sec. \ref{evaluation}, numerical results are presented. Sec. \ref{relatedwork} discusses related works. Finally, we conclude in Sec. \ref{con}.

\section{Content Dissemination with WiFi Direct}
\label{pro}

\subsection{WiFi Direct in Brief}
\label{WDoverview}

WiFi Direct is built on the prominent WiFi infrastructure mode \cite{conti2013experimenting}. It does not require dedicated hardware to support its functionalities. Therefore, it is now natively included in many mobile operating systems (e.g. Android $4.0$ and above). It enables devices to form groups for data exchange without the need of an AP. The topology of a group can be one-to-one or one-to-many. Within a group, a WiFi Direct device is selected to act as group owner (GO) to control the group including managing node join/leave, and starting/terminating the group. The GO is actually a soft AP that provides some functionalities of infrastructure AP, such as the basic service set (BSS) functionality, and WiFi Protected Setup \cite{alliance2014wi}. Other devices in this group, called clients, connect to the GO like connecting to an AP in a traditional WiFi network. To be the GO, a device has to be WiFi Direct enabled, while the clients can be WiFi Direct devices or normal WiFi devices. 

The roles of GO and client are not permanent. WiFi Direct specifies three modes of group formation, namely, standard, autonomous, and persistent. In the standard mode, to form a group, the devices need to negotiate and agree on the role that each device will act in this group. Technically, after these devices discover each other, they declare their desire to become the GO by sending a \textit{GO Intent} value to each other, and the device declaring the highest value becomes the GO \cite{camps2013device}. The process in the autonomous mode is much simpler, a device can autonomously create a group and select itself as the GO, and other devices that discover this group can join without negotiation. In the persistent mode, devices store network credentials and group information including GO and client roles for future usage. 

\subsection{GO-Coordinated Dissemination}

Content dissemination in MSNs exploits opportunistic contacts between mobile nodes. WiFi Direct is a favorable communication technology for such data dissemination due to its long transmission range and high data rate, in comparison to other alternatives such as Bluetooth and NFC\footnote{WiFi Direct supports typical WiFi speeds (maximum $250 Mb/s$) and a transmission range up to $200m$, whilst Bluetooth and NFC only support data rate up to $24 Mb/s$ and $424 kb/s$, and transmission range up to $100m$ and $0.2m$, respectively\cite{feng2014device}.}.

When a number of MSN nodes come into each other's transmission range, they first form a group by following one of the group formation processes of WiFi Direct. Once the group is established, the nodes can disseminate their data to other nodes in the group. WiFi Direct does not define the communication between clients \cite{FunaiTH16}, as each client does not know the information of other clients including IDs, MAC or IP addresses by its original design. Therefore, one has to implement additional function along with the MSN application to allow the data of all nodes being shared with others. To avoid changing the MAC and network layer of WiFi Direct, which may affect the operation of other WiFi Direct based applications, it is preferred to implement the additional function at the application layer. 

\begin{figure}[t]
\centering
\includegraphics[width=0.4\textwidth]{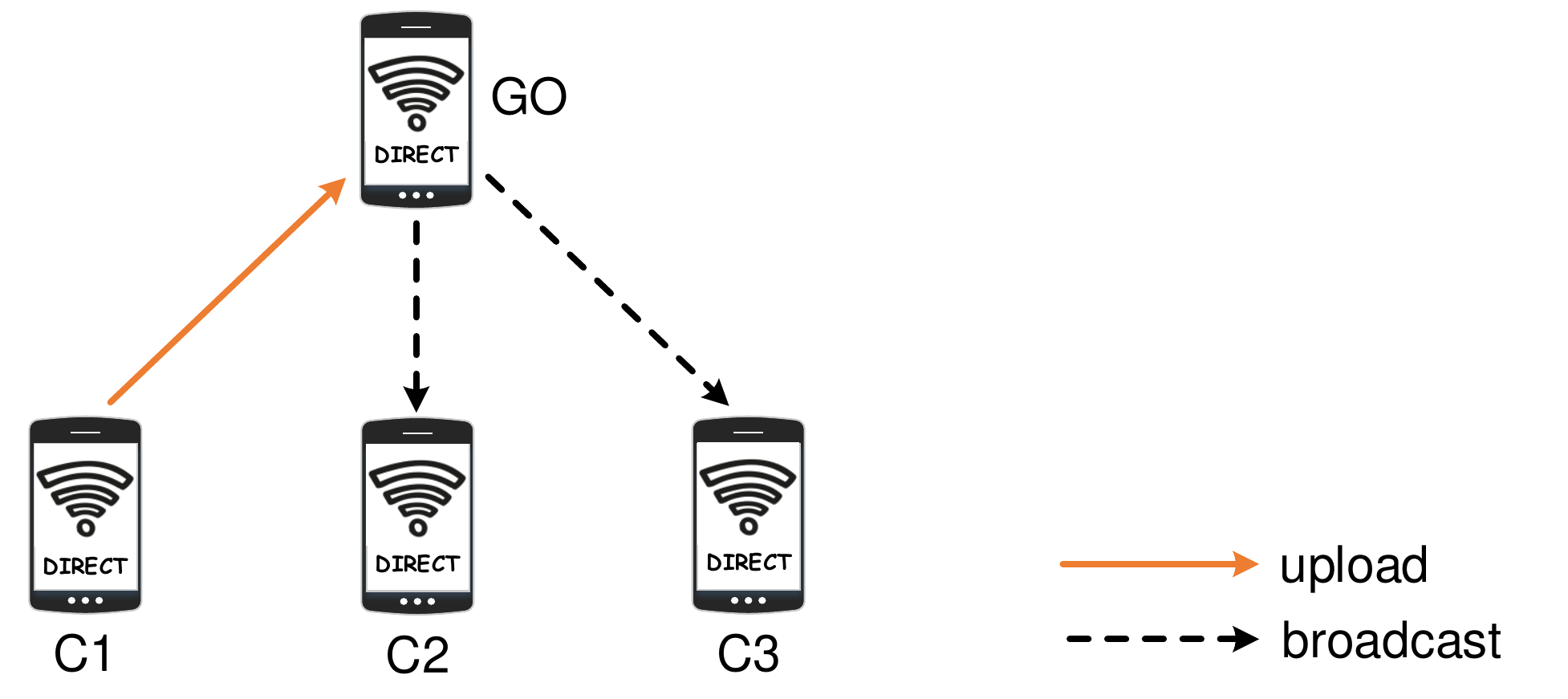}
\caption{An example of the GO-coordinated dissemination with a GO and three clients (C1, C2, and C3). The data of C1 is first uploaded to the GO and later broadcasted to other clients by the GO.}
\label{go-assist}
\end{figure}
Note that WiFi Direct is built on the WiFi infrastructure mode, all traffic between clients has to go through the GO\footnote{MAC layer broadcast does not go through the GO, however, it is not considered due to its unreliability \cite{chandra2009dircast}.}. Based on this feature, we propose a simple approach called \textit{GO-coordinated dissemination}. The basic idea is that, the clients upload their data to the GO that later broadcasts the received data for them (see Fig. \ref{go-assist} for an example). In addition, the GO allocates exclusive slots to every node (including the GO) and schedules all the data transmissions at the application layer. This can be realized simply by the GO sending the clients control messages to inform them to start/stop their transmissions. The point for such centralized scheduling is that WiFi Direct, like WiFi\footnote{Point coordination function is another MAC technique used in IEEE 802.11, which allows AP to coordinate the communication within the network, however it is not implemented by the Wi-Fi Alliance in its interoperability standard \cite{wiki-PCF}.}, uses distributed coordination function (DCF) to share the wireless channel among devices in the same group, and therefore nodes that have data to transmit need to content for channel access, which can cause severe collision and data retransmission when the data load is heavy. By centralized scheduling at the application layer, the GO-coordinated dissemination is able to alleviate channel contention.


%

\subsubsection*{The Two-Node Case}
\label{twonode}
When there are only two nodes in a group, they can transmit data to each other directly using unicast instead of the GO-coordinated dissemination. Once new nodes join the group, the GO-coordinated dissemination will be triggered. To this end, the GO checks the number of nodes in the group whenever a node joins or leaves the group, and selects proper transmission model accordingly.

\section{Fair Airtime Allocation Using Nash Bargaining}
\label{GNBSmodel}

In this section, we describe the fairness requirement in the GO-coordinated dissemination, formulate the airtime allocation for the GO-coordinated dissemination as a generalized Nash bargaining game, and analyze its solution that guarantees fair airtime allocation among nodes in a WiFi Direct group. 

\subsection{Fairness Requirement in GO-Coordinated Dissemination}

Consider a set $\mathcal{I}$ of nodes that have just formed a WiFi Direct group, $\mathcal{I}=\{1,2,...,I\}$.  Each node $i \in \mathcal{I}$ has a set of data, with total size $M_{i}$, to share with other nodes during this contact. Since MSNs typically employ store-carry-forward paradigm, the data to be shared can be readily determined by the network-level dissemination protocol (e.g. SSAR \cite{li2010routing} and PrefCast \cite{lin2012preference}) upon forming a group. In MSNs, nodes contact on the move, and therefore the contact duration can be so limited that some node(s) may not be able to finish disseminating all data. A study shows that the average contact duration of pedestrians with a mean speed of $1.3m/s$ is below $10$ seconds \cite{helgason2014opportunistic}. Though the GO-coordinated dissemination can alleviate the contention among the nodes regarding channel access, the nodes still have to compete for the limited airtime $T$ which is defined as the time available for data transmission during a contact. Therefore, it is significant to allocate the limited airtime to the nodes fairly.

Let $R^{b}$ be the broadcast data rate and $R^{u}_{i}$ be the uploading data rate of $i$ ($R^{u}_{i} = 0$ if $i$ is the GO since the GO does not upload data). Denote $(\mathbf{y};\mathbf{x}) = (y_{1},y_{2},...,y_{I};x_{1},x_{2},...,x_{I})$ an allocation of the limited airtime $T$, where $y_{i}$ is the allocated time to upload $i$'s data ($y_{i} = 0$ if $i$ is the GO) and $x_{i}$ is the allocated time to broadcast $i$'s data. In fact, we have $y_{i} = \frac{R^{b}}{R^{u}_{i}}\cdot x_{i}$ if we assume a stable loss rate during the whole contact. Then any feasible allocation $(\mathbf{y};\mathbf{x})$ is subject to the following constraints:
\begin{equation}
\sum_{i \in \mathcal{I}} (1+\beta_{i})x_{i} = T, 0\leq x_{i}\leq b_{i}, \forall i\in \mathcal{I}.
\end{equation}
where $\beta_{i} = \frac{R^{b}}{R^{u}_{i}}$ ($\beta_{i} = 0$ if $i$ is the GO) and $b_{i}$ is the estimated time required for the GO to broadcast all the data of $i$. Assuming no retransmission, then $b_{i}=\frac{M_{i}}{R}$.

Define \textit{content dissemination rate} $r_{k}$ of a given node $k$ the amount of $k$'s data per unit time received by all other nodes in the group. Then we have $r_{k} = \frac{R^{b}\cdot x_{k}}{T}$. In this paper, we aim for an allocation scheme that achieves fairness in content dissemination rate. To design such a scheme, the cooperative and self-interested behaviors of nodes have to be taken into consideration. On one hand, each node in MSNs benefits from the data dissemination, since it can receive data of its interests and its own data can be further disseminated by other nodes in the group in the future. On the other hand, nodes are effectively autonomous agents, since there is no network-wide control authority. Each node can decide, on its own will, whether to join the group and contribute resources to facilitate data dissemination. In addition, the node selected as the GO contributes more resources than client nodes. Therefore, it is reasonable to assume that each node seeks to maximize its utility from data dissemination over a contact. Such cooperative and self-interested nature of nodes makes this allocation problem perfectly fit into the analytical framework of generalized Nash bargaining game. Since the outcome of the bargaining game, which is called generalized Nash bargaining solution (GNBS), ensures Pareto optimality and achieves fairness in resource allocation, it is believed that GNBS is a suitable allocation policy in the context of local content dissemination in MSNs.

\subsection{Airtime Allocation Based on GNBS}

This section models the airtime allocation among nodes in a WiFi Direct group as a Nash bargaining game. In this game, players are the set $\mathcal{I}$ of nodes that are in contact and intend to share data through WiFi Direct, and the resource they bargain on is the limited airtime time $T$. Throughout bargaining, the players either reach an agreement on an airtime allocation, or come into disagreement. By the terminology of Nash bargaining theory, a possible allocation of transmission time is simply called a \textit{feasible agreement}. Denote $\mathcal{X} \subset \mathcal{R}^{I}$ the set of all possible agreements, $\mathbf{x} \in \mathcal{X}$, and $\mathbf{d}=(x^{d}_{1},x^{d}_{2},...,x^{d}_{I})$ the disagreement event. For each player $i \in \mathcal{I}$, there is a utility function $u_{i}(r_{i})$ that represents the degree of satisfaction for obtaining a dissemination rate of $r_{i}$. $u_{i}(r_{i})$ is assumed to be a differentiable, strict-increasing and concave function $\forall i\in \mathcal{I}$, meaning every node would like to obtain a high dissemination rate. Since $r_{i} = \frac{R^{b}\cdot x_{i}}{T}$, $u_{i}$ is a differentiable, strict-increasing and concave function of $x_{i}$ as well. Each feasible agreement in $\mathcal{X}$ results in a feasible utility vector $\mathbf{u}=(u_{1},u_{2},...,u_{I})$ in $\mathcal{U} \subset \mathcal{R}^{I}$, the set of all feasible utility vectors.

Formally, the Nash bargaining game is defined by the pair $(\mathcal{U},\mathbf{u^{d}})$ where $\mathbf{u^{d}}=(u_{1}(x^{d}_{1}),u_{2}(x^{d}_{2}),...,u_{I}(x^{d}_{I}))$ is the \textit{disagreement point}. The interpretation is that if no agreement is reached, then $i$ gets utility $u_{i}(x^{d}_{i}), \forall i\in \mathcal{I}$. Throughout, we assume that $\mathcal{U}$ is compact and convex, and there exists a $\mathbf{u} \in \mathcal{U}$ such that $u_{i} > u_{i}(x^{d}_{i}), \forall i \in \mathcal{I}$ which ensures that there exists a mutually beneficial agreement \cite{Osborne2005}.

Mathematically, GNBS, the optimal outcome of the generalized bargaining game, maximizes the following generalized Nash product (i.e. social welfare)
\begin{equation}\label{gnp1}
\begin{split}
\max_{\mathbf{x}} \textstyle\prod\limits_{i \in \mathcal{I}} (u_{i}(x_{i})-u_{i}(x^{d}_{i}))^{\alpha_{i}}, s.t.
\begin{cases} 
\sum\limits_{i \in \mathcal{I}} (1+\beta_{i})x_{i} = T \\
0\leq x_{i}\leq b_{i}, \forall i\in \mathcal{I}.
\end{cases}
\end{split}
\end{equation}
where $\alpha_{i}$ represents the bargaining power of player $i$, and $\sum^{I}_{i = 1} \alpha_{i} = 1$. The player with larger bargaining power could obtain higher dissemination rate and utility. In the content dissemination, the GO is entitled to obtain a larger dissemination rate and utility, as it contributes more resources (e.g. battery power) than clients. Therefore, we assign larger bargaining power to the GO than to the clients. Since the function of $log$ is concave and monotonic, the above generalized Nash bargaining problem is equivalent to (see proof in \cite{Yaiche2000})
\begin{equation}\label{gnp2}
\begin{split}
\max_{\mathbf{x}} \sum^{I}_{i = 1} \alpha_{i}\log(u_{i}(x_{i})-u_{i}(x^{d}_{i})), s.t.
\begin{cases} 
\sum_{i \in \mathcal{I}} (1+\beta_{i})x_{i} = T \\
0 \leq x_{i}\leq b_{i}, \forall i\in \mathcal{I}.
\end{cases}
\end{split}
\end{equation}
Let $L_{i}(x_{i}) = \frac{1+\beta_{i}}{\alpha_{i}} \cdot \frac{u_{i}(x_{i})-u_{i}(x^{d}_{i})}{u'_{i}(x_{i})}$, $F_{i}(\frac{1}{\lambda}) = \sum_{n=i}^{I}(1+\beta_{n}) \cdot L^{-1}_{n}(\frac{1}{\lambda})$, $i=1,2,...,I$. Without loss of generality, we assume the players are indexed such that $L_{1}(b_{1})<L_{2}(b_{2})<\cdots<L_{I}(b_{I})$. Then, we have the following theorem:
\begin{Theorem}\label{pro-gnbs}
There exists a unique agreement $\mathbf{x^{\star}}=(x^{\star}_{1},x^{\star}_{2},...,x^{\star}_{I})$ that induces the GNBS, which can be found by the following algorithm
\begin{equation}\label{wf-bridge}
 \left.\begin{aligned}
x^{\star}_{i}=\min \Bigg\{b_{i};L^{-1}_{i}\Bigg(F^{-1}_i\bigg(T-\sum^{i-1}_{j=1}(1+\beta_{j}) x^{\star}_{j}\bigg)\Bigg)\Bigg\},  \\
i=1,2,...,I.
\end{aligned}\right.
\end{equation}
\end{Theorem}

It is easy to find that the above algorithm has a computational complexity of $O(I)$ where $I$ is the number of nodes in the group. Since the constraints are linear, the objective function of problem (\ref{gnp2}) is a sum of concave functions, and hence concave, we are able to prove Theorem \ref{pro-gnbs} with Karush-Kuhn-Tucker (KKT) conditions \cite{boyd2004convex}. The complete proof is given in the Appendix. With $\mathbf{x^{\star}}$, the optimal allocation of broadcast time found by (\ref{wf-bridge}), the optimal allocation for uploading can be readily given by $\mathbf{y^{\star}} = (\beta_{1}x^{\star}_{1},\beta_{2}x^{\star}_{2},...,\beta_{I}x^{\star}_{I})$.

\subsubsection*{Allocation for The Two-Node Case}

Content dissemination for the two-node case does not need data uploading from the client to the GO. Letting $\mathcal{I}=\{1,2\}$ and $\beta_{1}, \beta_{2}=0$, the airtime allocation for the two-node case can also be modeled by the GNBS (\ref{gnp1}). In addition, the optimal allocation for the two-node case can be found by (\ref{wf-bridge}) as well.

\section{GNBS-based Scheduling Approach to Achieve Fair Allocation}
\label{sec-algorithm}

\begin{figure} 
\centering
\includegraphics[width=0.3\textwidth]{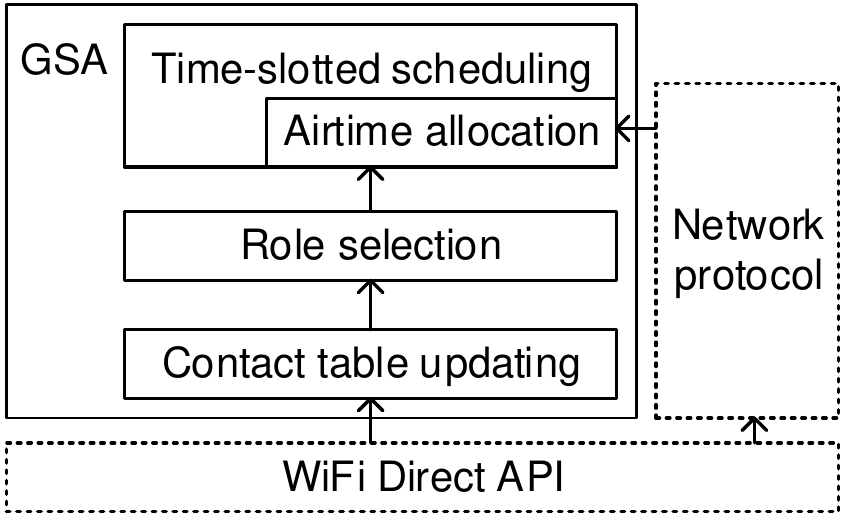}
\caption{Components of GSA.}
\label{fig-gsa}
\end{figure}

In this section, we present a GNBS-based scheduling approach (GSA) to achieve the fair allocation. The goal of GSA is three-fold: 1) to select a suitable GO that can make better use of the limited airtime; 2) to determine the allocation interval, namely, the length of airtime to be allocated; and 3) to schedule the transmissions (i.e. uploading and broadcast) of all the nodes. Fig. \ref{fig-gsa} shows the structure and components of GSA.

\subsection{Role Selection}
\label{goselection}

When several nodes come into contact, they first discover each other with the discovery service defined by WiFi Direct. After the discovery phase, each node summarizes how much data it wants to share, which is determined by the used routing protocol, and estimates how long it will stay in contact with other nodes. Then it sends a message containing information of its data load and the estimated contact duration to the others. Afterwards, they have to negotiate the roles of GO and client. 

We assume that the nodes are capable of estimating a \textit{pairwise contact duration} (PCD) with any other node, based either on their contact history or movements. For this, it has been shown by literature studies on contact traces that the pairwise contact duration of nodes in MSN-like networks follows certain distributions (e.g., power-law \cite{pietilanen2012dissemination}, log-normal \cite{helgason2014opportunistic}), and the nodes can use the mean value of the contact duration as the estimated contact duration. Alternatively, the nodes can compute an estimated contact duration with their mobility characteristics such as velocity and moving distance \cite{kim2014use}. Denote $d_{i}^{j}$ the estimated PCD between node $i$ and $j$. We assume $d_{i}^{j}=d_{j}^{i}$ for any pair of nodes. Upon joining the group, each node creates a \textit{contact table} that records the ID, PCD and total data size of all the nodes in contact. The table will be updated whenever a node leaves or a new node joins the group, and it will be deleted when the node itself leaves the group. Detailed contact table updating is described in Algorithm \ref{alg-ctable}. 

\begin{algorithm}
\caption{Contact table updating (executed by each $i$)}
\label{alg-ctable}
\begin{algorithmic}[1]
\Initialize {\textbf{create} a contact table upon joining a group}
\While {new node $k$ joins}
        \State add $<ID, PCD, M>_{k}$ 
        \State $\mathcal{I} \gets \mathcal{I}\cup\{k\}$
\EndWhile    
    \While {node $k$ leaves}
        \State  remove $<ID, PCD, M>_{k}$ 
        \State $\mathcal{I} \gets \mathcal{I}/\{k\}$
 \EndWhile    
    \While {it leaves}
        \State delete the contact table
\EndWhile
\end{algorithmic}
\end{algorithm}

\begin{figure}
    \centering
    \includegraphics[width=0.17\textwidth]{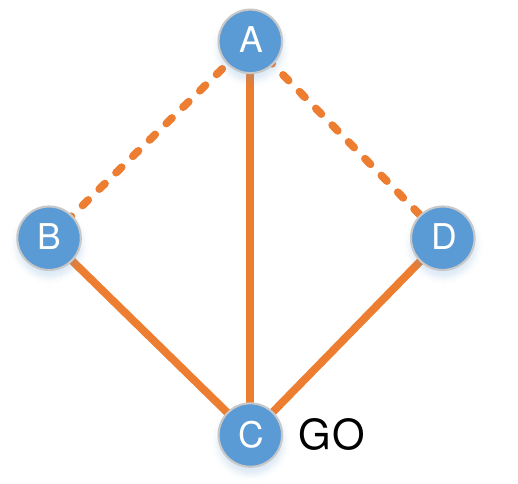}
       \quad \quad
       \scriptsize 
\begin{tabular}[b]{cc}
\hline
Node ID  & Load($ms$) \\ \hline
A & 10\\
B & 20\\
C & 30\\
D & 40\\ \hline
\vspace{3ex}
\end{tabular}
    \caption{An illustration of GO selection, where the solid lines are real connections after the group is formed. A and C are two candidates for the GO, since they both have direct connection with other nodes. Assume the data rate is $10mb/s$. Then, it needs $(10+2\times(20+30+40))/10=19s$ for all the nodes finish broadcasting their data if A is the GO, while it needs only $(30+2\times(10+20+40))/10=17s$ if C is the GO. Clearly, C is more suitable to be the GO.}
\label{fig-goselect}
\end{figure}

To be the GO, one node has to be able to build direct connections with all other nodes, so that every client is reachable via the GO. If there are multiple such nodes\footnote{If there is no such node, they can form multiple groups and negotiate for a dedicated channel for each group. However, that is out of our scope.}, then the one with the largest data load will be selected as the GO. In Fig. \ref{fig-goselect}, both node A and C can build direct connection with other nodes. Since having larger load than A, C will be selected as the GO. The above role selection scheme can be summarized as Algorithm \ref{alg-goselection}. It will be executed immediately after the contact table is updated. 

\begin{algorithm}
\caption{Role selection (executed by each node $i$)}
\label{alg-goselection}
\begin{algorithmic}[1]
\While {the contact table is updated}
\If {$i \in \mathcal{I'}$} \Comment $\mathcal{I'}$ is set of nodes that are able to build direct connections with all other nodes.
    \If {$i = \arg \max_{k} \{M_{k}, \forall k \in \mathcal{I'}\}$}
    \State set $role = GO$   
    \Else
    \State set $role = Client$
    \EndIf
        \Else
        \State set $role = Client$
\EndIf
\EndWhile
\end{algorithmic}
\end{algorithm}

\subsection{Allocation Interval}
\label{subsec-cde}

Normally, nodes in a group join or leave at different times due to their mobility. And any group change (e.g. node join and leave) necessarily triggers a new allocation among remaining group members. That basically means there would be many rounds of allocation during the lifetime of a group. Therefore, it is important to find the allocation interval $T$, the time for each round of allocation. We let $T= \min_{k}{d^{k}_{GO}}$, the shortest PCD between the GO and other nodes. If a larger interval is used, the node $i = \arg \min_{k}{d^{k}_{GO}}$ will not receive data from some other nodes and vice versa, since it is supposed to leave the group at $\min_{k}{d^{k}_{GO}}$. 

\begin{figure} 
\centering
\includegraphics[width=0.42\textwidth]{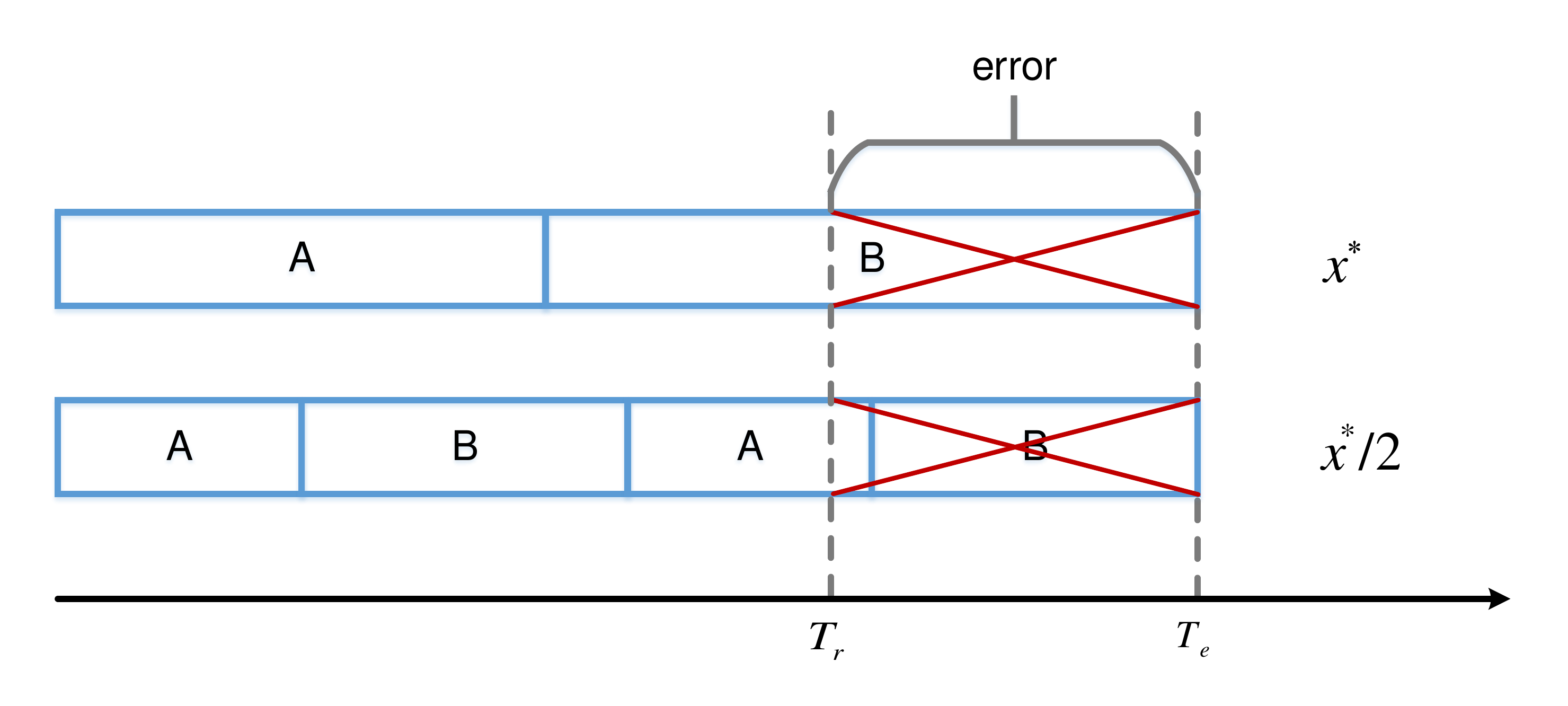}
\caption{Unfairness to node B caused by estimation error of contact duration. In the figure, $T_{r}$ is real contact duration, while $T_{e}$ is the estimated contact duration which is used for allocation.}
\label{fig-error}
\end{figure}

\subsection{Time-Slotted Scheduling}
\label{scheduling}

The allocation by GNBS relies on an estimation of the contact duration. The estimation error of contact duration would compromise the optimality of the allocation in terms of fairness (See Fig. \ref{fig-error} for an example). In order to reduce the unfairness caused by the estimation error, the allocated time by GNBS will be broken into small transmission slots. In addition, during each allocation interval, the transmission of all the nodes in $\mathcal{I}$ is scheduled in a round-robin way.

The slot for a given node $i$ is composed of two sub-slots, i.e., an \textit{uploading slot} and a \textit{broadcast slot}. During the uploading slot, node $i$ sends its data to the GO, while during the broadcast slot, the GO broadcasts the received data from $i$ to the other nodes. The whole slot size is given by 
\begin{equation}\label{window2}
W_{i} = \frac{(1+\beta_{i})x^{\star}_{i}}{\min_{k} \{(1+\beta_{k})x^{\star}_{k}\}}\cdot t_{slot}
\end{equation}
where $t_{slot}$, an engineering parameter, denotes the basic slot size. Then, the sizes of the uploading slot and the broadcast slot can be immediately obtained, which are
\begin{equation}\label{window21}
W^{u}_{i} = \beta_{i} \cdot W_{i}/(1+\beta_{i}) \text{ and } W^{b}_{i} = W_{i}/(1+\beta_{i}),
\end{equation}
respectively.  

\begin{figure}
  \centering
    \includegraphics[width=0.50\textwidth]{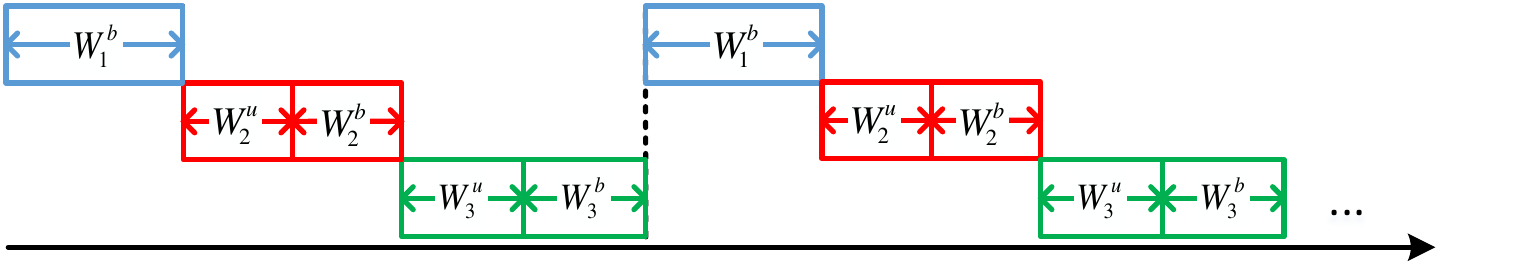}
  \caption{Time-slotted scheduling among three nodes for the GO-coordinated dissemination where node 1 is the GO.}
  \label{sche} 
\end{figure}

An example of the time-slotted scheduling is illustrated in Fig. \ref{sche}. The time-slotted scheduling is executed by the GO. To create a schedule, the GO needs the client to send their individual information to it, as specified in Algorithm \ref{wins-sch}. After the calculation, a schedule will be sent to each client. Finally, all the nodes transmit their data by following the schedule. 

\begin{algorithm}
\caption{Time-slotted scheduling (executed by the GO)}
\label{wins-sch}
\begin{algorithmic}[1]
\Require $T$ and $(u_{i}, x^{d}_{i}, b_{i}, \beta_{i})$ from each client
\While {all information received}
    \State \textbf{calculate} $(\mathbf{y^{\star}}, \mathbf{x^{\star}})$ using Eq. (\ref{wf-bridge})
    \State \textbf{calculate} $\mathbf{W^{u}} = (W^{u}_{1},W^{u}_{2},...,W^{u}_{I})$ and $\mathbf{W^{b}} = (W^{b}_{1},W^{b}_{2},...,W^{b}_{I})$ using Eq. (\ref{window2}) and (\ref{window21})
	\State \textbf{send} $\mathbf{W^{u}}$, $\mathbf{W^{b}}$ and $\mathbf{t_{start}}$ to each client
\EndWhile
\end{algorithmic}
\end{algorithm}

\section{Numerical Study}
\label{evaluation}

In this section, we consider a basic system setup and evaluate the performance of GSA through numerical study. We assume the loss probability is uniformly distributed in $[0,0.1]$. Low loss is assumed due to little contention on channel access among nodes in the group. The estimation error of contact duration follows a normal distribution $N(0,1)$. Uploading rate and broadcast rate are both set to $11mb/s$. Default basic slot size $t_{slot}$ is set to $20ms$. For the utility function, we use the following normalized form:
\begin{equation}
u_{i} = \frac{r_{i}}{r^{max}_{i}} =\frac{ \frac{R^{b}\cdot x_{i}}{T}}{ \frac{R^{b}\cdot b_{i}}{T}} = \frac{x_{i}}{b_{i}}
\end{equation}
where $u_{i} \in [0,1]$. We assign the same bargaining power $\alpha_{c}$ to all clients, and a bargaining power $\alpha_{g} = 2\alpha_{c}$ to the GO. Lastly, the disagreement point $\mathbf{u^{d}}$ is set to $\mathbf{0}$ without loss of generality. In our future work, we will study different forms of utility function and disagreement point under a more general system setup. 

\subsection{Fairness in Airtime Allocation}

We consider a WiFi Direct group $\mathcal{I}$ comprising $6$ nodes, $\mathcal{I}=\{n1,n2,n3,n4,n5,n6\}$. Their data loads are $[10,20,40,40,60,80]$ (in $mb$). The following two schemes are used to compare with our GSA: 
\begin{itemize}
\item \textit{Equal allocation (EQL)}. The broadcast slot sizes of all nodes are equal. 
\item \textit{Weighted allocation (WTD)}. The broadcast slot sizes of all nodes are proportional to their requirements. 
\end{itemize}

\begin{table}[!t]
\caption{Allocated uploading/broadcast time $y_{i}/x_{i}$ ($s$). }
\label{tab-allocation}
\centering
\resizebox{0.45\textwidth}{!} {
\begin{tabular}{c|cc|cc|cc}
     \hline
\multirow{2}{*}{ Node $i$} & \multicolumn{2}{c|}{GSA} & \multicolumn{2}{c|}{EQL} & \multicolumn{2}{c}{WTD} \\
\cline{2-7}
              & $y_{i}$  &  $x_{i}$              & $y_{i}$  &  $x_{i}$               & $y_{i}$  &  $x_{i}$     \\ 
\hline
     $n1$ &   0.714 & 0.714     & 0.909 & 0.909      &  0.217  &  0.217  \\
     $n2$ &   0.714 & 0.714     & 0.909 & 0.909      &  0.435  &  0.435  \\
     $n3$ &   0.714 & 0.714     & 0.909 & 0.909      &  0.869  &  0.869  \\
     $n4$ &   0     & 2.857     & 0     & 0.909      &  0      &  0.869  \\
     $n5$ &   0.714 & 0.714     & 0.909 & 0.909      &  1.304  &  1.304  \\
     $n6$ &   0.714 & 0.714     & 0.909 & 0.909      &  1.739  &  1.739  \\          
     \hline
     \end{tabular}
}
\end{table}

\begin{figure} 
\centering
\includegraphics[width=0.415\textwidth]{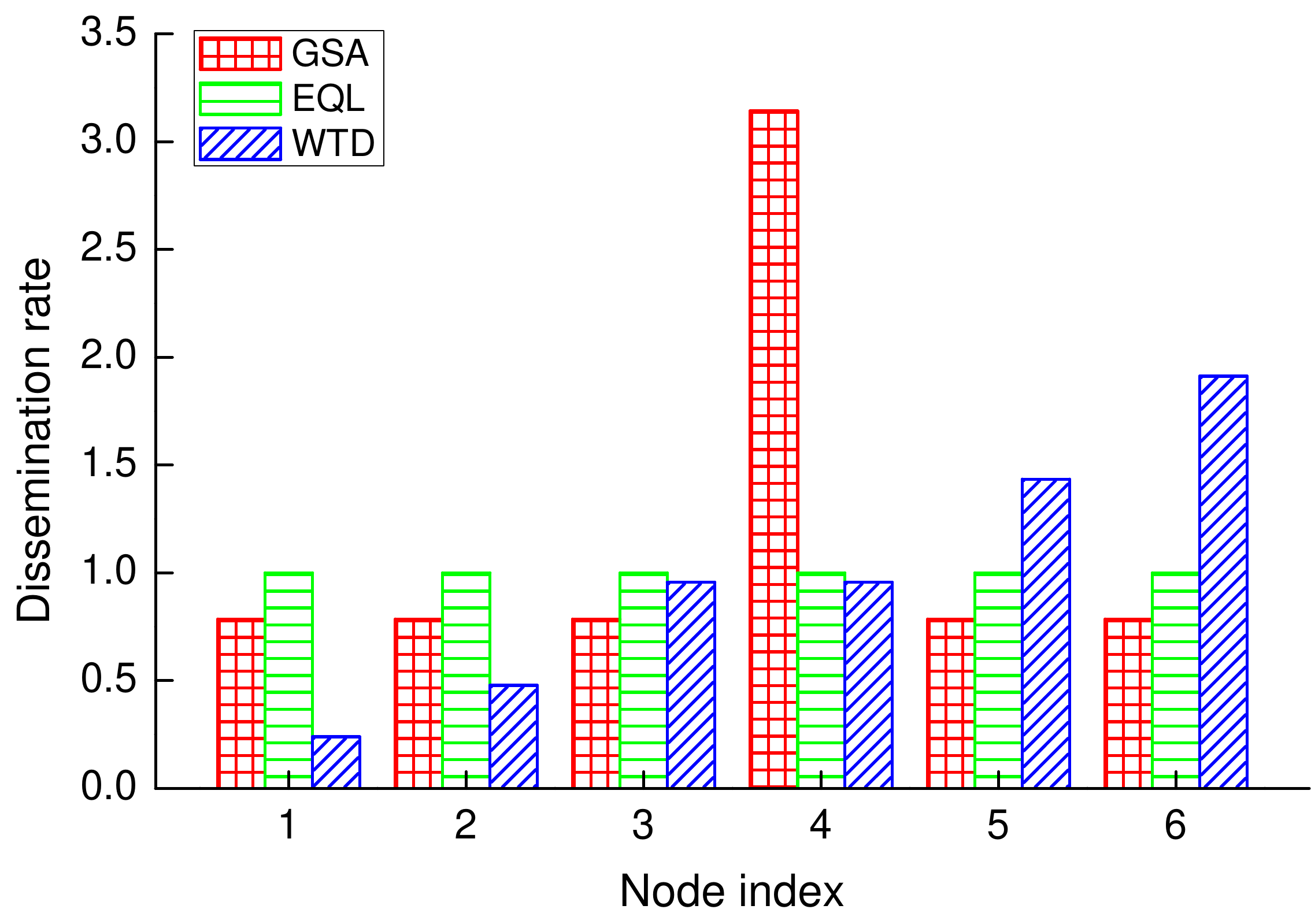} 
\caption{Comparison of GSA, EQL and WTD when $ T = 10s$ .}
\label{all-gnbs}
\end{figure}

Table \ref{tab-allocation} shows the allocation results of GSA, EQL and WTD for an instance with allocation interval $T = 10s$ and $n4$ acting as the GO. As the GO, $n4$ does not need to spend time on uploading. GSA allocates equal broadcast time to all client nodes while allocates a notably larger amount of time to the GO (i.e. $n4$). Fig. \ref{all-gnbs} illustrates the resulting dissemination rates of the nodes. It can be seen that clients obtain equal dissemination rate, while the GO gets a much larger rate due to its larger bargaining power. It indicates that GSA provides fairness in dissemination rate while capturing the asymmetric contributions of nodes.

\begin{figure} 
\centering
\includegraphics[width=0.4\textwidth]{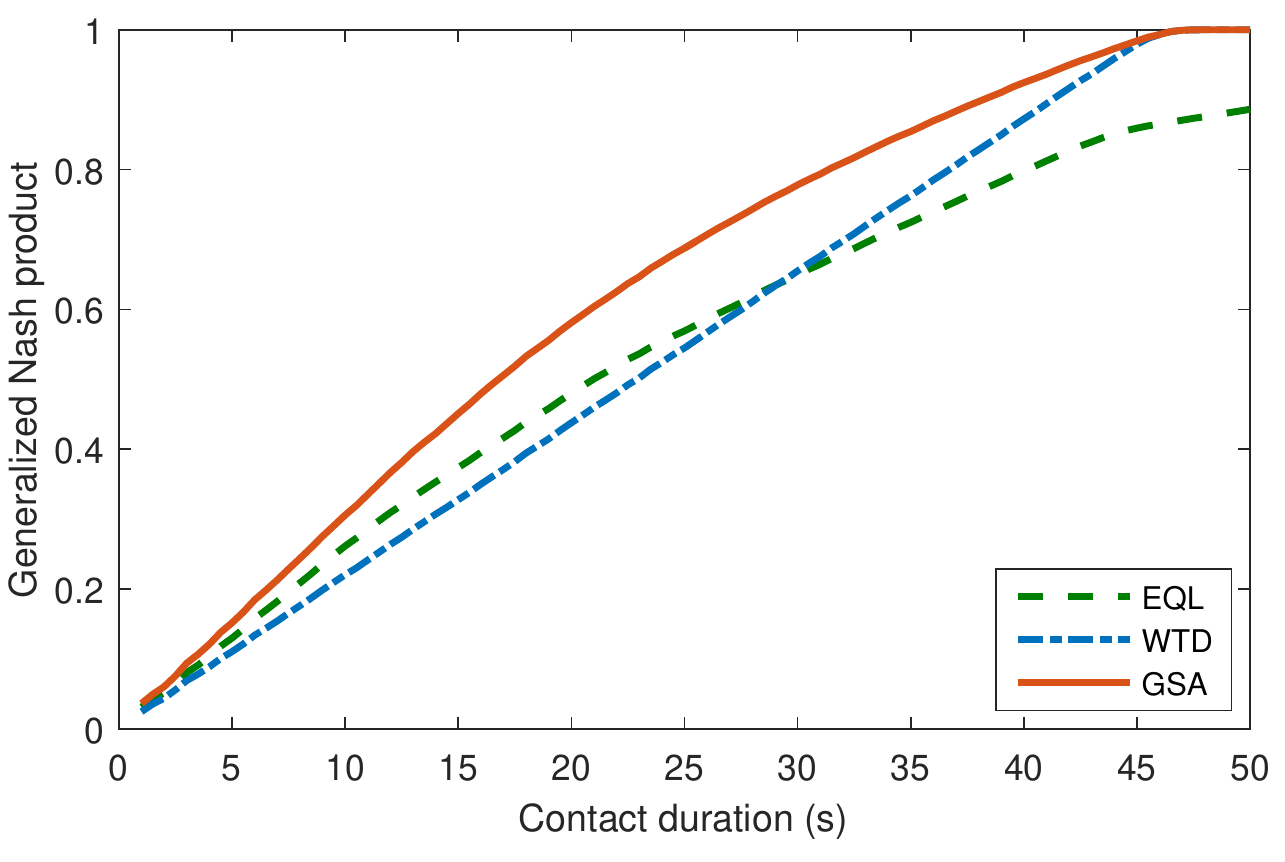} 
\caption{Generalized Nash products of GSA, EQL and WTD over different contact durations.}
\label{gnbs-wpf}
\end{figure}

\begin{figure} 
\centering
\includegraphics[width=0.425\textwidth]{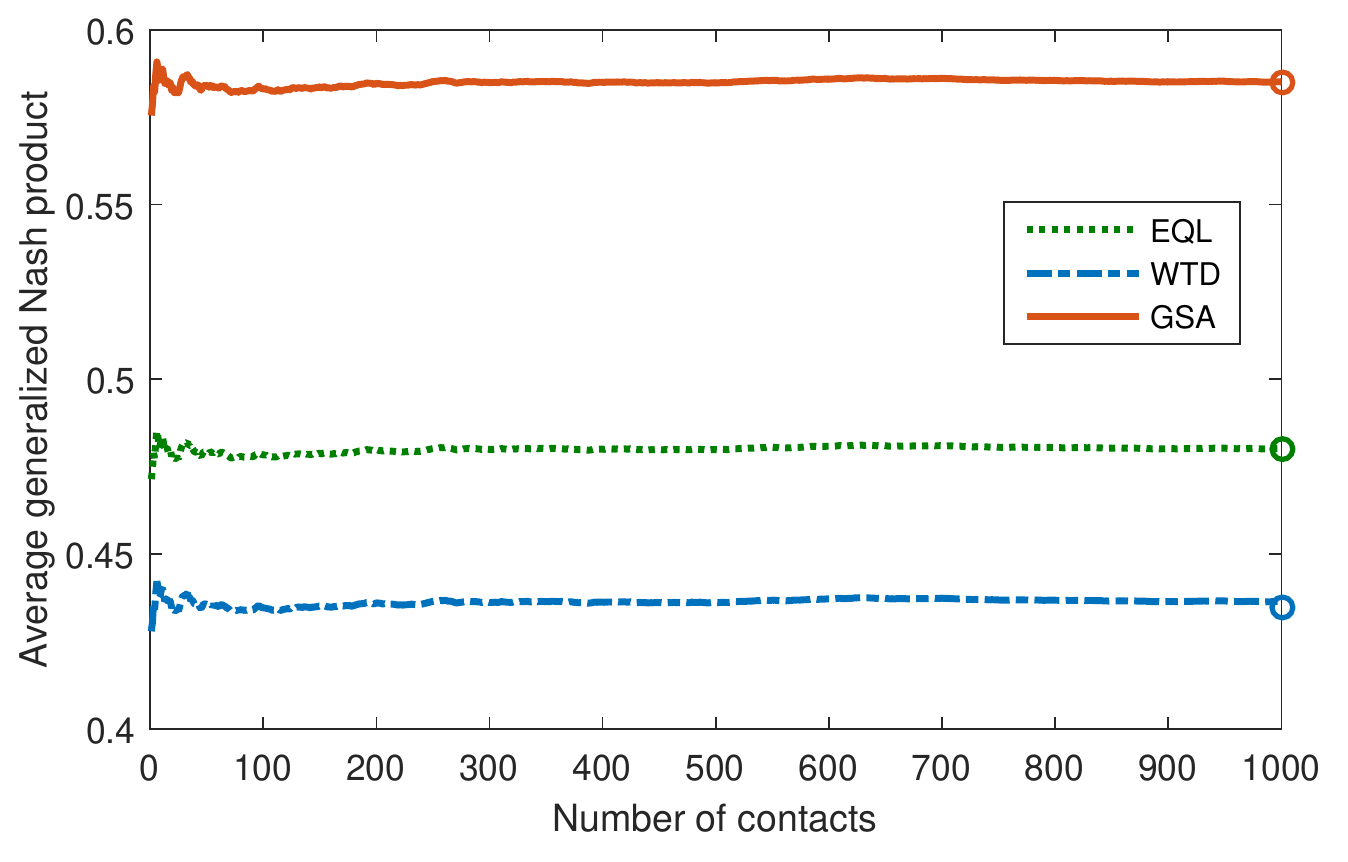}
\caption{Average generalized Nash product over time. The circle at the end of each curve denotes the maximum generalized Nash product without randomness.} 
\label{gnp-vstime}
\end{figure}

It is easy to show that GNBS guarantees weighted proportional fairness in utility \cite{kelly1998rate} when $\mathbf{u^{d}}=\mathbf{0}$. It means that moving away from the GNBS point $\mathbf{u}^{GNBS}$ to another point $\mathbf{\bar{u}} \in \mathcal{U}$ will not increase the aggregate of weighted proportional changes in utilities: 
\begin{equation}\label{eq-wpf}
\sum^{I}_{i=1} \alpha_{i}\cdot \frac{\bar{u}_{i}-u_{i}^{GNBS}}{u_{i}^{GNBS}} \leq 0.
\end{equation} 
Fig. \ref{gnbs-wpf} shows a comparison between GSA, EQL and WTD in terms of weighted proportional fairness. Each point represents an average of $1000$ runs, reflecting the randomness of the contact duration. As expected, GSA always has larger generalized Nash product than EQL and WTD. Fig. \ref{gnp-vstime} illustrates that the average generalized Nash product for a specific mean contact duration, i.e., $20s$, versus the number of contacts of this group of nodes who are likely to meet with each other continually over time. The average generalized Nash product for the first few contacts fluctuates, due to impact of the estimation error. However, it does not take too many times of contact to converge to the theoretical maximum generalized Nash product. 

\begin{figure}
\centering
\includegraphics[width=0.42\textwidth]{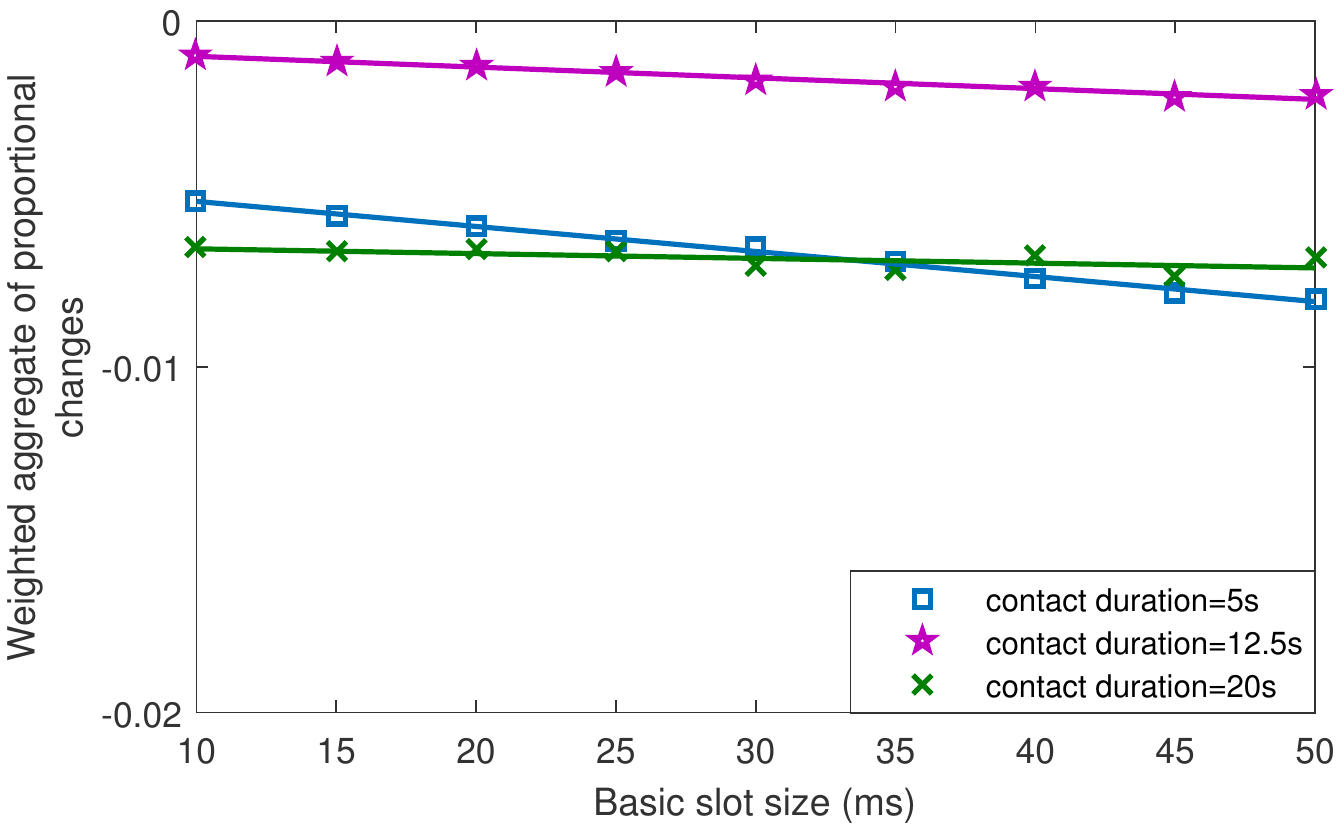}
\caption{Achieved weighted proportional fairness by the time-slotted scheduling of GSA with different basic slot sizes.}
\label{agg-gnbs}
\end{figure}

The GNBS based airtime allocation used by GSA relies on a contact duration estimation. In practice, the estimation may hardly achieve perfect accuracy. As a result, the weighted proportional fairness of GSA could be compromised. Fig. \ref{agg-gnbs} illustrates the aggregate of weighted proportional changes over different basic slot sizes. As can be seen, the aggregate is slightly below zero, and decreases almost linearly with the basic slot size. To achieve better fairness, small basic slot size is preferred.

\subsection{Dynamic Join/Leave of Nodes} 
\begin{figure}
\centering
\includegraphics[width=0.45\textwidth]{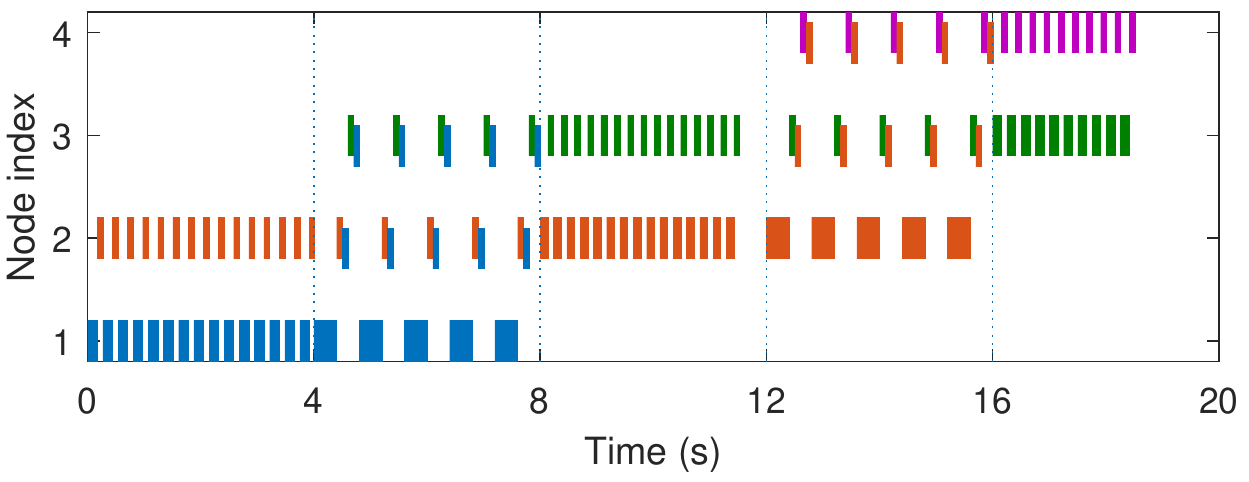}
\caption{The schedule for $\{n1,n2,n3,n4\}$ during $(0,20]s$.}
\label{fig-sche2}
\end{figure}

In this simulation, we show the adaptivity of GSA to dynamic join/leave of nodes into the group. Consider four nodes $\{n1,n2,n3,n4\}$ that join the group at $[0,0,4,12]s$ and leave the group at $[8,16,20,20]s$. They have $[25, 20, 15, 10]mb$ data for each of the rest nodes. Since each node join or leave triggers a new round of allocation, there will be five rounds of allocation, and the allocation intervals are all $4$ seconds. The basic slot size is set to $100ms$. Fig. \ref{fig-sche2} shows the schedule for the four nodes. It can be seen that when there are three nodes in the group (i.e., during $(4,8]s$ and $(12,16]s$), the GO relays data for the clients. Each client uploads its data to the GO during its uploading slots, followed by the GO broadcasting the data to other clients. GSA rewards the GO with much higher broadcast slot size than the clients, as can be noted in Table \ref{t-wins2}. 

\begin{table}[!t]
\caption{Uploading/broadcast slot sizes $W_{u}/W_{b}$ ($ms$). }
\label{t-wins2}
\centering
\resizebox{0.49\textwidth}{!} {
\begin{tabular}{c|rr|rr|rr|rr|rr}
     \hline
\multirow{2}{*}{ Node} & \multicolumn{2}{c|}{$(0-4]s$} & \multicolumn{2}{c|}{$(4-8]s$} & \multicolumn{2}{c|}{$(8-12]s$} & \multicolumn{2}{c|}{$(12-16]s$} & \multicolumn{2}{c}{$(16-20]s$} \\
\cline{2-11}
              & $W_{u}$  &  $W_{b}$              & $W_{u}$  &  $W_{b}$               & $W_{u}$  &  $W_{b}$                & $W_{u}$  &  $W_{b}$                  & $W_{u}$  &  $W_{b}$  \\ \hline
     $n1$ &                 & 167                         &                 & \cellcolor{lightgray}400                        &                 &                               &                  &                                &                 & \\
     $n2$ &                 & 100                         & 100          & 100                        &                 & 133                          &                  &\cellcolor{lightgray} 400                           &                 & \\
     $n3$ &                 &                                & 100          & 100                        &                 & 100                          & 100           & 100                           &                 & 150\\
     $n4$ &                 &                                &                 &                             &                 &                               & 100           & 100                           &                 & 100\\
     \hline
     \end{tabular}
}
\end{table}

\section{Related Work}
\label{relatedwork}

\subsection{Contact Duration in MSNs}

In MSNs, contact duration is normally short due to the mobility and short transmission range of nodes. As a result, the capacity of a contact is limited. There are a number of recent works considering the significance of limited contact duration in their protocol design \cite{zhao2016contact,zhuo2011social,li2015coff,kim2014use,li2015contact,jung2013link,trifunovic2013slicing,chaves2013improving}, though it tends to be overlooked by earlier works on routing, forwarding in MSNs. LCD \cite{jung2013link} improves the shortest-path routing by diverting some message(s) to other nodes not on the shortest path if a node is on the shortest paths of too many messages. Large content normally needs long duration to propagate. Given that the contact duration follows an exponential law distribution, Y. Li et al. show that content size is one of the most important factors to influence the content dissemination delay\cite{li2015contact}. Assuming that data of large size could not be completely transmitted during a single contact, authors propose fragmentation-based schemes for data replication\cite{zhao2016contact}, cooperative caching\cite{zhuo2011social}, traffic offloading\cite{li2015coff}, and message forwarding \cite{kim2014use}, respectively. These schemes generally divide each data piece into a number of small packets or blocks at the source node with techniques such as network coding, and send a subset of them to each node that comes into contact.

The above works assume that nodes contact with each other in a pairwise manner, which has been a predominant assumption in most MSN literature. However, in \cite{wennerstrom2015considering}, simultaneous multiple contact among nodes is found to be quite common in real-world contact traces. Therefore, group communication can be more efficient than pairwise communication for content dissemination if multiple nodes are in contact at the same time. In this paper, we follow up this idea and propose GO-coordinated dissemination for MSN content dissemination. More differently from the above works, we study the contact duration from a resource allocation's perspective. 

\subsection{Application of Nash Bargaining}
Nash bargaining solution has been extensively used to model resource allocation problems in computer networks, such as bandwidth allocation\cite{Yaiche2000} and Internet access sharing \cite{iosifidis2014enabling}. As an extension of NBS, generalized Nash bargaining solution, also called asymmetric Nash bargaining solution, is able to capture the differences among players in bargaining power. H. Boche et al. use GNBS to study resource allocation among users with different priorities in wireless networks\cite{boche2007non}. In \cite{park2007bargaining,wang2014generalized}, the authors apply GNBS to the problem of rate allocation among multimedia users with a distortion-based utility function. To provide fairness, different bargaining powers are assigned to users with heterogeneous video characteristics. Based on a GNBS modeling, Falloc algorithm provides minimum bandwidth guarantee while allocates residual bandwidth fairly among all virtual machine pairs in a data center\cite{guo2016fair}. In this paper, we use GNBS to model the airtime allocation problem among a group of mobile nodes in WiFi-Direct-based MSNs.

\section{Conclusions}
\label{con}
In this paper, we studied local content dissemination in a WiFi-Direct-based MSN. Specifically, we proposed an intuitive GO-coordinated dissemination strategy that does not require change on the WiFi Direct protocol. We designed a Nash bargaining based fair airtime allocation to decide how long each node can use to transmit data during the limited contact duration. Since the optimal allocation given by the bargaining model cannot be directly implemented due to that the estimation of the contact duration may be inaccurate, we designed a time-slotted scheduling approach that divides the allocated time into smaller slots and allows nodes to transmit in a slot at a time. Finally, we validated the designed allocation scheme and scheduling approach through numerical study. 

\section*{Acknowledgment}
The work was partly supported by the EU FP7 CLIMBER project (www.fp7-climber.eu).

\appendix[Proof of Theorem \ref{pro-gnbs}]

The proof consists of two steps: 1) we prove that there is a unique agreement $\mathbf{x^{\star}}$ induces the GNBS by using Lemma \ref{lemmaIIA} - \ref{pie-func}; and 2) we prove that the unique agreement can be found by (\ref{wf-bridge}) with Lemma \ref{app-lemma}. 

\begin{Lemma}\label{lemmaIIA}
Node $i$ will not consider alternatives in $[0,x^{d}_{i}]$, for any $i \in \mathcal{I}$.
\end{Lemma}
\begin{IEEEproof}
Equivalently, this lemma states that, if $\mathbf{x^{\star}}$ exists, then $x^{\star}_{i}>x^{d}_{i}$. Suppose $x^{\star}_{i} \leq x^{d}_{i}$. Since $u_{i}(x_{i})$ is a strict-increasing function, it is easy to see that $u_{i}(x^{\star}_{i}) \leq u_{i}({x}^{d}_{i})$ holds and the objective function $P(\mathbf{x})=\max_{\mathbf{x}} \sum^{I}_{i = 1} \log(u_{i}(x_{i})-u_{i}(x^{d}_{i}))$ in (\ref{gnp2}) is increasing in $\mathbf{x}$. Since it has been assumed that there exists a mutually beneficial agreement, we have $\exists {u}_{i}(\tilde{x}_{i}) > u_{i}({x}^{d}_{i}) \geq u_{i}(x^{\star}_{i})$ for every $i$. As a result, $P(\mathbf{\tilde{x}})>P(\mathbf{x^{\star}})$. It apparently contradicts the fact that $\mathbf{x^{\star}}$ maximizes $P(\mathbf{x})$. Therefore, player $i$ will not consider alternatives in $[0,x^{d}_{i}]$, $ \forall i \in \mathcal{I}$.
\end{IEEEproof}
 
Lemma \ref{lemmaIIA} helps us reduce the constraint of $0\leq x_{i}\leq b_{i}$ to ${x}^{d}_{i} < x_{i}\leq b_{i}$. Consequently, the optimization problem (\ref{gnp2}) is equivalent to
\begin{equation}\label{snb2}
\begin{split}
\max_{\mathbf{x}} \sum^{I}_{i = 1} \alpha_{i}\log(u_{i}(x_{i})-u_{i}(x^{d}_{i})), s.t.
\begin{cases} 
\sum_{i \in \mathcal{I}} (1+\beta_{i})x_{i} = T \\
x^{d}_{i}< x_{i}\leq b_{i}, \forall i\in \mathcal{I}.
\end{cases}
\end{split}
\end{equation}

Since the constraints are linear, the objective function of problem (\ref{snb2}) is a sum of concave functions, and hence concave, we can find the unique optimal solution with KKT conditions \cite{boyd2004convex}. The Lagrangian is
\begin{eqnarray}
\ell & = & \sum_{i=1}^{I} \alpha_{i} \log(u_{i}(x_{i})-u_{i}(x^{d}_{i}))-\lambda(\sum_{i=1}^{I}(1+\beta_{i})x_{i}-T) \nonumber\\   
 & &-\sum_{i=1}^{I}\mu_{i}(x^{d}_{i}-x_{i})-\sum_{i=1}^{I}\nu_{i}(x_{i}-b_{i})
\end{eqnarray}
where $\lambda\geq 0$, $\mu_{i}\geq 0$ and $\nu_{i}\geq 0$, $\forall i \in \mathcal{I}$ are Lagrangian multipliers. The optimality conditions are given by
\[ \begin{split}
\frac{\alpha_{i} u'_{i}(x_{i})}{u_{i}(x_{i})-u_{i}(x^{d}_{i})}-\lambda(1+\beta_{i})+\mu_{i}-\nu_{i}=0\\
\sum_{i=1}^{I} (1+\beta_{i}) x_{i}-T=0 \\
\mu_{i}(x^{d}_{i}-x_{i})=0 \\
\nu_{i}(x_{i}-b_{i})=0 & \quad\quad\quad \forall i \in \mathcal{I}.
\end{split} \]

\begin{Lemma}\label{inc-func}
Let $L_{i}(x_{i}) = \frac{1+\beta_{i}}{\alpha_{i}} \cdot \frac{u_{i}(x_{i})-u_{i}(x^{d}_{i})}{u'_{i}(x_{i})}$. $L_{i}(x_{i})$ is an increasing function on $ (x^{d}_{i},b_{i}]$.
\end{Lemma}
\begin{IEEEproof}
For each node $i$, the utility function $u_{i}(x_{i})$ is a differentiable, strict-increasing and concave function. It implies that $u'_{i}(x_{i})>0$ and $u''_{i}(x_{i})<0$ where $u'_{i}(x_{i})$ and $u''_{i}(x_{i})$ are the first and the second derivatives of $u_{i}(x_{i})$ with respect to $x_{i}$, respectively. By $u''_{i}(x_{i})<0$, we know that $u'_{i}(x_{i})$ decreases with $x_{i}$. Given any $x^{1}_{i}, x^{2}_{i} \in (x^{d}_{i},b_{i}]$, if $x^{1}_{i} > x^{2}_{i}$, then $u_{i}(x^{1}_{i}) > u_{i}(x^{2}_{i})>u_{i}(x^{d}_{i})$ and $u'_{i}(x^{2}_{i})>u'_{i}(x^{1}_{i})$. Then, it is easy to see that $\frac{1+\beta_{i}}{\alpha_{i}} \cdot \frac{u_{i}(x^{1}_{i})-u_{i}(x^{d}_{i})}{u'_{i}(x^{1}_{i})}>\frac{1+\beta_{i}}{\alpha_{i}} \cdot \frac{u_{i}(x^{2}_{i})-u_{i}(x^{d}_{i})}{u'_{i}(x^{2}_{i})}>0$. Hence, $L_{i}(x_{i})$ strictly increases with $x_{i} \in (x^{d}_{i},b_{i}]$. Moreover, $L^{-1}_{i}(\cdot)$ is a strict-increasing function on $(L_{i}(x^{d}_{i}),L_{i}({b}_{i})]$. 
\end{IEEEproof}

Now we start to solve these equations, to find $\mathbf{x}$, $\lambda$, $\mathbf{\mu}$, and $\mathbf{\nu}$. First of all, it can be easily seen that $\mu_{i}=0$ for all $i \in \mathcal{I}$, since $x^{d}_{i} < x_{i}$. Note that $\nu_{i}$ acts as a slack variable in the last equation, so it can be eliminated, leaving
\begin{eqnarray}
\lambda \leq \frac{1}{L_{i}(x_{i})}  \label{kkt1} \\
\sum_{i=1}^{I} (1+\beta_{i}) x_{i}-T=0 \\
(\frac{1}{L_{i}(x_{i})}-\lambda)(x_{i}-b_{i})=0 & \quad \forall i \in \mathcal{I}.  \label{kkt3}
\end{eqnarray}
By Lemma \ref{inc-func}, we know that $\frac{1}{L_{i}(x_{i})}$ strictly decreases with $x_{i}$ on $ (x^{d}_{i},b_{i}]$. Therefore, 1) if $\lambda^{\star} > \frac{1}{L_{i}(b_{i})}$, (\ref{kkt1}) holds only if $x_{i} < b_{i}$. It implies that $\lambda^{\star} = \frac{1}{L_{i}(x_{i})}$, by (\ref{kkt3}). Then we have $x^{\star}_{i} = L^{-1}_{i}(\frac{1}{\lambda^{\star}})$ if $\lambda^{\star} > \frac{1}{L_{i}(b_{i})}$; 2) if $\lambda^{\star} < \frac{1}{L_{i}(b_{i})}$, which with $\frac{1}{L_{i}(b_{i})} \leq \frac{1}{L_{i}(x_{i})}$ implies $\lambda^{\star} < \frac{1}{L_{i}(x_{i})}$, we get $x^{\star}_{i} = b_{i}$, by (\ref{kkt3}); and 3) if $\lambda^{\star} = \frac{1}{L_{i}(b_{i})}$, (\ref{kkt3}) becomes $(\frac{1}{L_{i}(x_{i})}-\frac{1}{L_{i}(b_{i})})(x_{i}-b_{i})=0$, which has a unique solution $x^{\star}_{i} = b_{i} = L^{-1}_{i}(\frac{1}{\lambda^{\star}})$. In conclusion, we have 
\begin{equation}\label{unique-solution0}
\begin{split}
x^{\star}_{i}=
\begin{cases} 
L^{-1}_{i}(\frac{1}{\lambda^{\star}}), & \quad \text{if~} \frac{1}{\lambda^{\star}} \leq L_{i}(b_{i}) \\
b_{i}, & \quad \text{if~} \frac{1}{\lambda^{\star}} > L_{i}(b_{i}), \quad \forall i \in \mathcal{I}.
\end{cases}
\end{split}
\end{equation}
More simply, 
\begin{equation}\label{unique-solution}
x^{\star}_{i}=\min \{b_{i};L^{-1}_{i}(\frac{1}{\lambda^{\star}})\}, \forall i \in \mathcal{I}.
\end{equation}
Substituting (\ref{unique-solution}) into the condition $\sum_{i=1}^{I} (1+\beta_{i}) x_{i}-T=0$, we obtain 
\begin{equation}\label{uniqie-lambda}
\sum_{i=1}^{I} (1+\beta_{i}) \cdot \min \{b_{i};L^{-1}_{i}(\frac{1}{\lambda^{\star}})\}=T.
\end{equation}

\begin{figure} 
\centering
\includegraphics[width=0.34\textwidth]{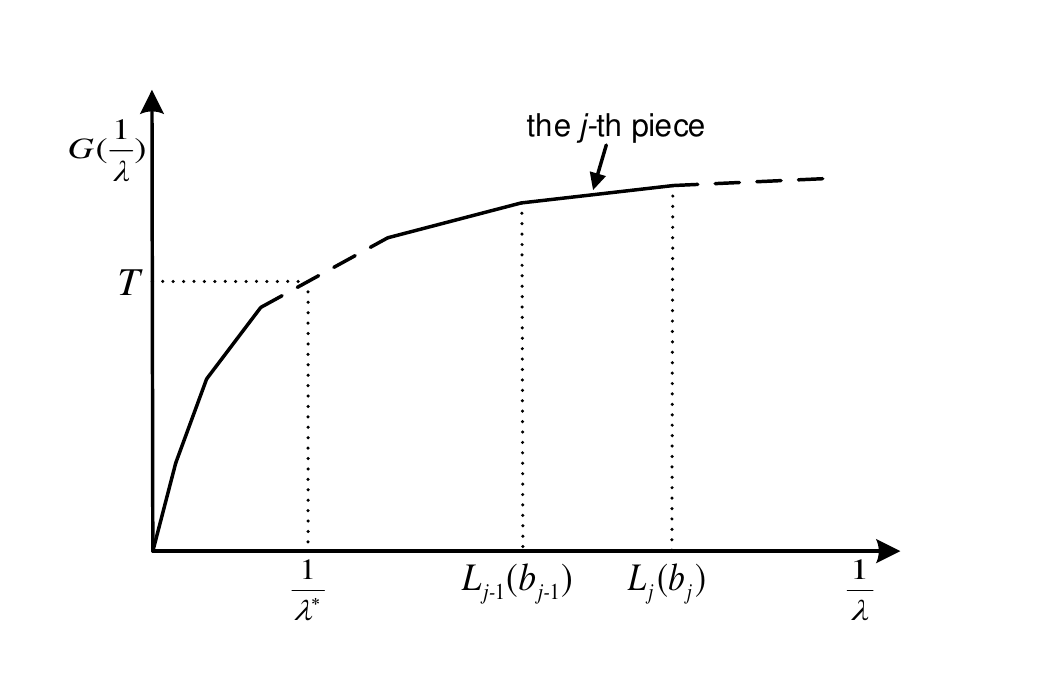}
\caption{$G(\frac{1}{\lambda})$. The gradient at $\frac{1}{\lambda}$ on each piece $j$ (not including its start and end points) is given by $\frac{d(\sum_{n=j}^{I}(1+\beta_{n})L^{-1}_{n}(\frac{1}{\lambda}))}{d(\frac{1}{\lambda})}$.}
\label{lambda}
\end{figure}

Without loss of generality, we assume the players are indexed such that $L_{1}(b_{1})<L_{2}(b_{2})<\cdots<L_{I}(b_{I})$. 

\begin{Lemma}\label{pie-func}
Let $G(\frac{1}{\lambda}) = \sum_{i=1}^{I} (1+\beta_{i}) \cdot \min \{b_{i};L^{-1}_{i}(\frac{1}{\lambda})\}$. Then $G(\frac{1}{\lambda})$ is a strict-increasing function on $(0,L_{I}(b_{I})]$. 
\end{Lemma}

\begin{IEEEproof}
As depicted in Fig. \ref{lambda}, $G$ is a piecewise function of $\frac{1}{\lambda}$, with breakpoints at $L_{i}(b_{i})$, $i = 1,2,...,I$. For the $j$th piece defined on $(L_{j-1}(b_{j-1}),L_{j}(b_{j})]$, $G_{j}(\frac{1}{\lambda})=\sum_{m=1}^{j-1}  (1+\beta_{m}) b_{m}+F_{j}(\frac{1}{\lambda})$ where $F_{i}(\frac{1}{\lambda}) = \sum_{n=i}^{I}(1+\beta_{n}) \cdot L^{-1}_{n}(\frac{1}{\lambda})$, $j=1,2,...,I$. In the proof of Lemma \ref{inc-func}, we have shown that $L^{-1}_{j}(\cdot)$ is a strict-increasing function on $(0,L_{j}(b_{j})]$, $j = 1,2,...,I$. Evidently, $F_{j}(\frac{1}{\lambda})$ and $G_{j}(\frac{1}{\lambda})$ strictly increases with $\frac{1}{\lambda}$. Hence, $G(\frac{1}{\lambda})$ in overall strictly increases with $\frac{1}{\lambda}$ on $(0,L_{I}(b_{I})]$. 
\end{IEEEproof}

It is easy to verify that $0<\frac{1}{\lambda^{\star}}< L_{I}(b_{I})$. Otherwise, we get $\sum_{i=1}^{I} x^{\star}_{i}=0$ if $\frac{1}{\lambda^{\star}}=0$, and $\sum_{i=1}^{I} (1+\beta_{i}) x^{\star}_{i}=\sum_{i=1}^{I} (1+\beta_{i}) b_{i}>T$ if $\frac{1}{\lambda^{\star}}\geq L_{I}(b_{I})$, which contradict $\sum_{i} (1+\beta_{i}) x^{\star}_{i} = T$. Since $G(\frac{1}{\lambda}) = \sum_{i=1}^{I}(1+\beta_{i}) \min \{b_{i};L^{-1}_{i}(\frac{1}{\lambda})\}$ strictly increases with $\frac{1}{\lambda}$ on $(0,L_{I}(b_{I})]$, stated in Lemma \ref{pie-func}, there is only one $\frac{1}{\lambda^{\star}}$ that results in $\sum_{i=1}^{I} (1+\beta_{i}) \min \{b_{i};L^{-1}_{i}(\frac{1}{\lambda^{\star}})\}=T$, namely, (\ref{uniqie-lambda}). With unique $\frac{1}{\lambda^{\star}}$, we can argue that $x^{\star}_{i}$ in (\ref{unique-solution}) is unique as well.

The above proves that there is a unique solution to the bargaining game. In the following, we prove that the unique solution can be given by (\ref{wf-bridge}). To avoid confusion, denote $\bar{\mathbf{x}}=(\bar{x}_{1},\bar{x}_{2},...,\bar{x}_{I})$ the agreement found by (\ref{wf-bridge}). It has been proved that there is a unique agreement $\mathbf{x^{\star}}=(x^{\star}_{1},x^{\star}_{2},...,x^{\star}_{I})$ that satisfies (\ref{unique-solution}). Therefore, to prove that (\ref{wf-bridge}) can find the unique agreement, we can show that $\bar{x}_{i}=x^{\star}_{i}$, namely,
\begin{equation*}
\resizebox{.49 \textwidth}{!} 
{$\min \big\{b_{i};L^{-1}_{i}\big(F^{-1}_i\big(T-\sum^{i-1}_{j=1}(1+\beta_{j}) x^{\star}_{j}\big)\big)\big\}=\min \{b_{i};L^{-1}_{i}(\frac{1}{\lambda^{\star}})\}$}
\end{equation*}
for every $i=1,2,...,I$. First of all, let us introduce the following lemma. 
\begin{Lemma}\label{app-lemma}
If $\bar{x}_{j}=x^{\star}_{j}$ for $j=1,...,i-1$, then $\bar{x}_{i}=x^{\star}_{i}$.
\end{Lemma}
\begin{IEEEproof}
Let $\frac{1}{\lambda_{i}}=F^{-1}_{i}(T-\sum^{i-1}_{j=1}(1+\beta_{j}) \bar{x}_{j})$. Then we have $\bar{x}_{i}=\min \{b_{i};L^{-1}_{i}(\frac{1}{\lambda_{i}})\}$. In the following, we show that $\bar{x}_{i}=x^{\star}_{i}$ for the cases of $\frac{1}{\lambda^{\star}} \geq L_{i}(b_{i})$ and $\frac{1}{\lambda^{\star}}<L_{i}(b_{i})$.

1) If $\frac{1}{\lambda^{\star}} \geq L_{i}(b_{i})$, we can prove that $\frac{1}{\lambda_{i}} \geq L_{i}(b_{i})$ by contradiction. Assume to the contrary that $\frac{1}{\lambda_{i}} < L_{i}(b_{i})$. Since $L^{-1}_{j}(\cdot)$ is an increasing function for every $j$, we have 
\begin{equation}\label{appenb}
b_{i}=L^{-1}_{i}(L_{i}(b_{i}))>L^{-1}_{i}(\frac{1}{\lambda_{i}}),
\end{equation}
and
\begin{equation}\label{appentwo2}
L^{-1}_{k}(\frac{1}{\lambda^{\star}})>L^{-1}_{k}(\frac{1}{\lambda_{i}}), k = i+1,...,I.
\end{equation}
Further, since $ L_{i}(b_{i}) < L_{k}(b_{k})$, we have 
\begin{equation}\label{appentwo1}
b_{k}=L^{-1}_{k}(L_{k}(b_{k}))>L^{-1}_{k}(L_{i}(b_{i}))>L^{-1}_{k}(\frac{1}{\lambda_{i}}).
\end{equation}
From (\ref{appentwo2}) and (\ref{appentwo1}), it can be seen that
\begin{equation}\label{appenmin}
\min \{b_{k};L^{-1}_{k}(\frac{1}{\lambda^{\star}})\}>L^{-1}_{k}(\frac{1}{\lambda_{i}}), k=i+1,...,I. 
\end{equation}
With $\frac{1}{\lambda^{\star}} \geq L_{i}(b_{i})$, from (\ref{unique-solution}) we know that $x^{\star}_{i}=b_{i}$. Then, by (\ref{uniqie-lambda}), we have
\begin{equation}\label{appenT}
\begin{split}
T =& \sum_{j=1}^{i-1} (1+\beta_{j})x^{\star}_{j}+(1+\beta_{i})b_{i}+  \\
 & \sum_{k=i+1}^{I} (1+\beta_{k}) \min \{b_{k};L^{-1}_{k}(\frac{1}{\lambda^{\star}})\}.
\end{split}
\end{equation}
Using the assumption that $\bar{x}_{j}=x^{\star}_{j}$ for $j=1,...,i-1$,  and the inequalities in (\ref{appenb}) and (\ref{appenmin}) gives
\begin{eqnarray}\label{appenT2}
T  &  > & \sum_{j=1}^{i-1} (1+\beta_{j})\bar{x}_{j}+(1+\beta_{i}) L^{-1}_{i}(\frac{1}{\lambda_{i}})+ \nonumber\\   
& & \sum_{k=i+1}^{I} (1+\beta_{k}) L^{-1}_{k}(\frac{1}{\lambda_{i}}) \nonumber\\
 & = & \sum_{j=1}^{i-1} (1+\beta_{j}) \bar{x}_{j}+F_{i}(\frac{1}{\lambda_{i}}).
\end{eqnarray}
Because $\frac{1}{\lambda_{i}}=F^{-1}_{i}(T-\sum^{i-1}_{j=1}(1+\beta_{j}) \bar{x}_{j})$, the right side of (\ref{appenT2}) equals $T$. Now we have a contradiction. Therefore, $\frac{1}{\lambda_{i}} \geq L_{i}(b_{i})$ holds. As a result, $\bar{x}_{i}=b_{i}=x^{\star}_{i}$.

2) If $\frac{1}{\lambda^{\star}} < L_{i}(b_{i})$, we prove that $\frac{1}{\lambda_{i}} = \frac{1}{\lambda^{\star}}$. By $\frac{1}{\lambda^{\star}} < L_{i}(b_{i})$, we have $x^{\star}_{k}=L^{-1}_{k}(\frac{1}{\lambda^{\star}})$, for $k=i,...,I$. Then it follows from (\ref{uniqie-lambda}) that $T=\sum_{j=1}^{i-1} (1+\beta_{j}) x^{\star}_{j}+\sum_{k=i}^{I} (1+\beta_{k}) L^{-1}_{k}(\frac{1}{\lambda^{\star}})$. Equivalently, 
\begin{equation}\label{appenT5}
\sum_{k=i}^{I}(1+\beta_{k}) L^{-1}_{k}(\frac{1}{\lambda^{\star}})  =  T -\sum_{j=1}^{i-1} (1+\beta_{j}) x^{\star}_{j}.
\end{equation}
The left part of (\ref{appenT5}) is effectively $F_{i}(\frac{1}{\lambda^{\star}})$, while the right part equals $F_{i}(\frac{1}{\lambda_{i}})$ because $T -\sum_{j=1}^{i-1} (1+\beta_{j}) x^{\star}_{j}=T -\sum_{j=1}^{i-1} (1+\beta_{j}) \bar{x}_{j}=F_{i}(\frac{1}{\lambda_{i}})$ under the assumption that $\bar{x}_{j}=x^{\star}_{j}$ for $j=1,...,i-1$. In other words, $F_{i}(\frac{1}{\lambda^{\star}})=F_{i}(\frac{1}{\lambda_{i}})$which implies $\frac{1}{\lambda_{i}} = \frac{1}{\lambda^{\star}}$. Consequently, $\bar{x}_{i}=x^{\star}_{i}$. This completes the proof of the lemma.
\end{IEEEproof}

It remains to show that $\bar{x}_{1}=x^{\star}_{1}$. In fact, it can be easily proved by letting $i=1$ and following exactly the same steps with the proof of Lemma \ref{app-lemma}. With $\bar{x}_{1}=x^{\star}_{1}$, we can get that $\bar{x}_{i}=x^{\star}_{i}$, $i=2,3,...,I$ by iteratively applying Lemma \ref{app-lemma}.


\end{document}